\documentclass[apj,preprint]{aastex61}
\usepackage{graphicx}
\usepackage{natbib}
\usepackage{float}
\shorttitle{Pre-merger electromagnetic signals of binary compact stars}
\shortauthors{Wang, Peng, Wu, \& Dai}

\begin{document}

\title{Pre-merger electromagnetic counterparts of binary compact stars}

\author{Jie-Shuang Wang}

\affil{School of Astronomy and Space Science, Nanjing University, Nanjing 210093, China;
jiesh.wang@gmail.com,dzg@nju.edu.cn}
\affil{Tsung-Dao Lee Institute, Shanghai Jiao Tong University, Shanghai 200240, China}
\author{Fang-Kun Peng}
\affil{Guizhou Provincial Key Laboratory of Radio Astronomy and Data Processing, Guizhou Normal University, Guiyang 550001, China}
\affil{School of Physics and Electronic Science, Guizhou Normal University, Guiyang 550001, China}
\author{Kinwah Wu}
\affil{Mullard Space Science Laboratory, University College London, Holmbury St. Mary, Dorking,
Surrey, RH5 6NT, UK; kinwah.wu@ucl.ac.uk}
\author{Zi-Gao Dai}
\affil{School of Astronomy and Space Science, Nanjing University, Nanjing 210093, China;
jiesh.wang@gmail.com,dzg@nju.edu.cn}
\affiliation{Key Laboratory of Modern Astronomy and Astrophysics (Nanjing University), Ministry of Education, China}

\newcommand{\be}{\begin{equation}}
\newcommand{\ee}{\end{equation}}
\newcommand{\g}{\gamma}
\def\ba{\begin{eqnarray}}
\def\ea{\end{eqnarray}}
\def\cE{{\cal E}}
\def\cR{{\cal R}}
\def\bmu{{\mbox{\boldmath $\mu$}}}
\def\bphi{{\mbox{\boldmath $\phi$}}}

\begin{abstract}
We investigate emission signatures of binary compact star gravitational wave sources consisting of strongly magnetized neutron stars (NSs) and/or white dwarfs (WDs) in their late-time inspiral phase. Because of electromagnetic interactions between the magnetospheres of the two compact stars, a substantial amount of energy will be extracted, and the resultant power is expected to be $\sim 10^{38}-10^{44}$~erg/s in the last few seconds before the two stars merge, when the binary system contains a NS with a surface magnetic field $10^{12}$~G. The induced electric field in the process can accelerate charged particles up to the EeV energy range. Synchrotron radiation is emitted from energetic electrons, with radiative energies reaching the GeV energy for binary NSs and the MeV energy for NS-WD or double WD binaries. In addition, a blackbody component is also presented and it peaks at several to hundreds keV for binary NSs and at several keV for NS-WD or double WD binaries. The strong angular dependence of the synchrotron radiation and the isotropic nature of the blackbody radiation lead to distinguishable modulation patterns between the two emission components. If coherent curvature radiation is presented, fast radio bursts could be produced. These components provide unique simultaneous electromagnetic signatures as precursors of gravitational wave events associated with magnetized compact star mergers and short gamma ray bursts (e.g., GRB~100717). 
\end{abstract}

\section{Introduction}

The gravitational wave (GW) event GW170817 was recently detected
  by the advanced LIGO and Virgo \citep{Abbott2017a},
  and in less than 2~s an associated short gamma-ray burst (GRB~170817A) was observed
  \citep{Abbott2017b,Goldstein2017,Savchenko2017,Zhang2017,Li2018},
  confirming that $\gamma$-ray bursts can indeed be produced by neutron star (NS) mergers.
Subsequently, multiband electromagnetic (EM) counterparts were observed,
  respectively in the radio \citep{Hallinan2017,Alexander2017},
  optical \citep[dominated by the `kilonova', e.g.][]{Coulter2017,Abbott2017c},
  and X-ray \citep{Troja2017,Margutti2017} bands, respectively.
A summary of the multi-messenger and multiband follow-up observations
  of GW170817 was presented in \cite{Abbott2017c}.
This was the first detection of both a GW event and its EM counterparts.
These observations not only confirm the prediction that short GRBs originate from NS mergers
  but also provide strong support to the NS-black hole (BH) merger scenario
  \citep[see][]{Goodman1986,Paczynski1986,Eichler1989,Narayan1992}.

EM counterparts of GW events from mergers are of great importance in astrophysics and fundamental physics.
They provide a means of investigating the multi-facet, such as the dynamical behaviour
   of the GW sources throughout the entire merging process.
With NS mergers now established as GW sources,
  pre-merger EM counterparts of GW events would naturally be precursors of short GRBs.
Currently, much attention has been drawn to EM emission properties of the GW sources
  shortly preceding merger/coalescence events
  \citep[for reviews of EM counterparts of double NS and NS-BH mergers, see e.g.][]{Fernandez2016ARNPS}.
Previous research on the pre-merger EM counterparts
   has invoked the unipolar inductor (UI) process, operating in close binary systems
   where one of the compact star is strongly magnetized, extending its magnetic filed lines
   to the weakly magnetized (or non-magnetic) companion
   \citep[][see also \cite{Vietri1996,Hansen2001}]{McWilliams2011,Piro2012,Lai2012,Wang2016,D'Orazio2016}.
This UI model was analogous to the UI model
   originally proposed for the the Jupiter-Io system
   \citep[][see also \cite{Hess2007}]{Piddington1968,Goldreich1969},
   which was also generalized for white dwarf (WD) binaries
   \citep{Wu2002,Willes2004,Dall'Osso2006,Dall'Osso2007,Wu2008,Wu2009},
   planet-WD systems \citep{Li1998,Willes2004,Willes2005},
   the exo-planet-magnetic stars systems \citep{Zarka2001,Zarka2007,Laine2012},
   and the pulsar-planet system \citep{Mottez2011b,Mottez2011a,Mottez2014,Dai2016}.

\cite{Wang2016} proposed that fast radio bursts (FRBs) are produced
  via a UI process, expected to occur during the late-time inspiral of two NSs.
They analyzed the required conditions for successful FRBs and
  found that the EM energy-loss rate in the UI model
  is consistent with that of the EM power of NS-BH binary and double NS binary
  (for a ratio of 100 between the magnetic dipole moments of the two NSs)
  during their late-time inspiral phase
  derived from general relativistic magnetohydrodynamics (GRMHD) simulations
  \citep[see][]{Paschalidis2013,Palenzuela2013b,Ponce2014,Paschalidis2017}.
For merging of two NSs with comparable magnetic fields,
  GRMHD simulations showed that the radiative power generated
  could be much higher than that predicted by a simple UI process \cite[see][]{Palenzuela2013b,Ponce2014}.
Note that the interactions between two magnetic dipoles
  and the EM energy loss rate were also investigated by \cite{Ioka2000}.
However, as will be shown below, their assumption of a rigid dipole field is not unquestionable
 when the magnetic interaction deforms the structure of the magnetic fields in the late inspiralling phase.

In this paper, we investigate the EM emission of two compact stars during their late-time inspiral phase,
  due to the interaction between their magnetospheres.
The paper is organized as follows.
In the next section,
  we first investigate the time-dependent properties
  of EM energy loss from a binary compact star system.
The rate of EM energy loss had been studied previously in the GRMHD simulations.
However, the simulations were restricted to the last several milliseconds
  before the merging occurs \citep{Paschalidis2013,Palenzuela2013b,Ponce2014}.
Here, we adopt an analytic approach,
   which allows us to extend the time coverage
   and obtain an understanding of how the behavior of the EM counterpart
   proceeds to the final merging phase.
We presents theoretical multiband signals
  expected from these close orbiting compact binary systems in section 3
  and observational signatures, in particular spectral properties, in section 4.
A summary and discussion are given in section 5.

\section{Electromagnetic energy-dissipation rate}

We first calculate the EM radiation from double NS binaries during their inspiral.
Then we extend the calculations
  to the NS-WD binaries and double WD binaries.
The UI process in the double BH systems requires a different treatment
  and deserves a more thorough separate study \citep[e.g.][]{Zhang2016}.
The double BH systems are therefore not considered in this work.

Consider that a double NS binary
  comprises of a magnetic main star with a magnetic dipole moment $\mu_*=B_*R_*^3$
  and a companion with $\mu_{\rm c}=B_{\rm c}R_{\rm c}^3$,
  where $B_*$ is the characteristic surface magnetic field and $R_*$ is the radius of the main NS.
Here and after we use the subscript `$*$' to
  represent the main star, and `c' to represent the companion.
The binary is separated by a distance $a$ and orbits at an angular speed
   $\Omega=[GM_*(1+q)/a^3]^{1/2}$, where $M$ is the mass and $q=M_{\rm c}/M_*$.
We consider three cases, according to the magnetic fields of two NSs:
  case 0 with $B_{\rm c}<\mu_*a^{-3}$ (including the case of NS-BH binary),
  case 1 with $\bmu_*\sim- \bmu_{\rm c}$,
  and case 2 with $\bmu_*\sim\bmu_{\rm c}$.

\begin{figure}
\centering
\includegraphics[width=0.5\textwidth]{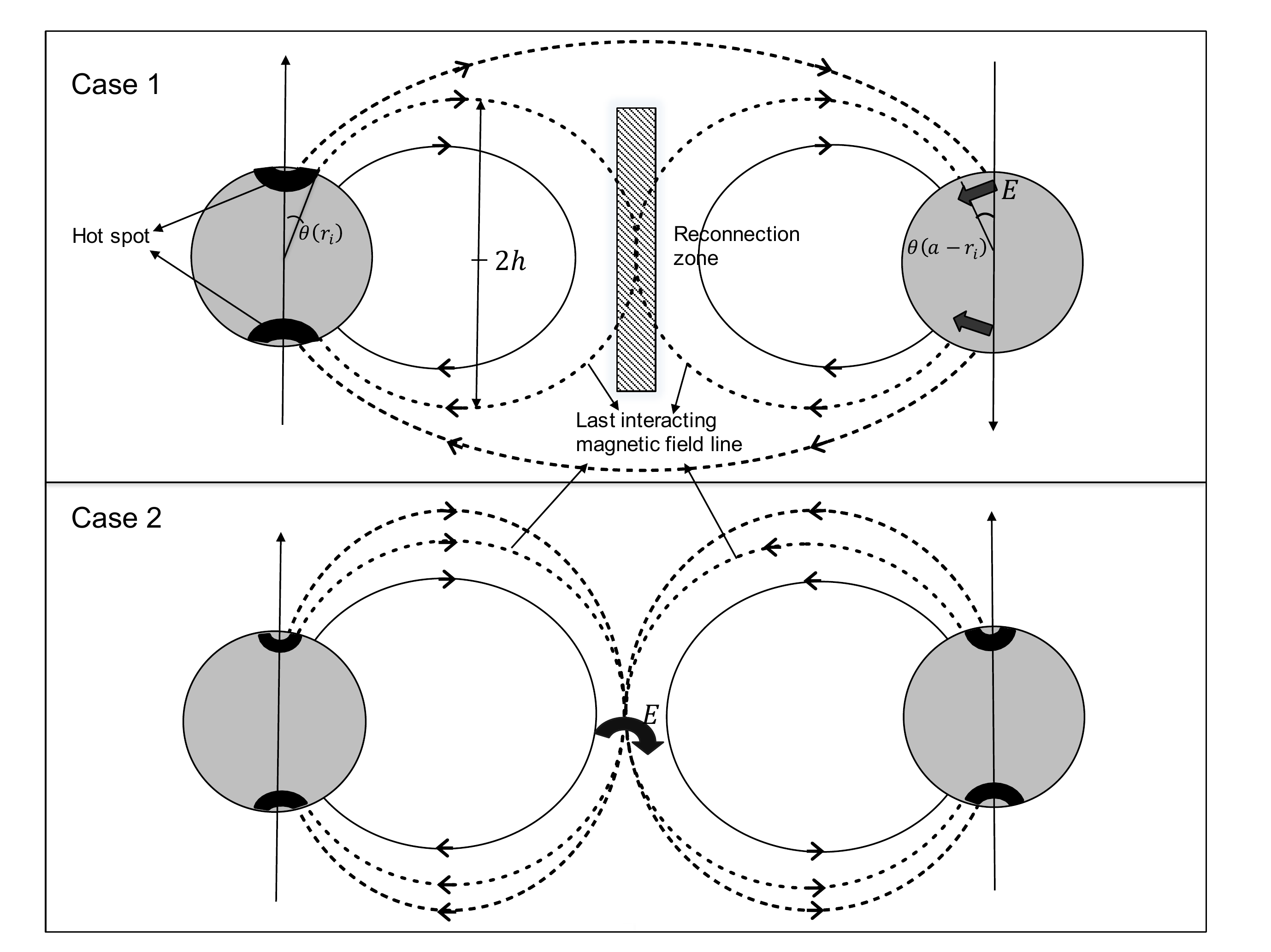}
\caption{Schematic pictures of the cases with $\bmu_*=\pm \bmu_{\rm c}$. The dashed
lines are the interaction magnetic field lines while the solid line is non-interacting. We use black
regions to mark hot spots and thick black arrows to mark an electric field. \label{fig1}}
\end{figure}

In case 0, a UI model is usually adopted.
The maximum EM energy-dissipation rate is then
\be
L_{\rm UI}\approx 1.7\times 10^{42} M_{*,1.4}(1+q)\mu_{*,30}^2 (R_{\rm c}/10\;\!{\rm km})^2 (a/ 30\;\!{\rm km})^{-7} {\rm erg/s},
\label{eqUI}
\ee
under the assumption that the resistance of the system is dominated by the magnetosphere \citep{Lai2012},
  where $ M_{*,1.4}$ is the main-star mass, in units of $1.4M_\odot$,
  and $\mu_{*,30}=\mu_*/10^{30}\,{\rm G\;\!cm}^3$.
The energy-dissipation rate of the NS-BH binary can be obtained
  by replacing the $R_{\rm c}$ with the Schwarzschild radius of the BH. 
However, when $B_{\rm c}>\mu_*a^{-3}$, the magnetic interaction also depends on magnetic configurations of both NSs.
We study two typical cases as examples: one is that the magnetic dipole moments of the NSs are anti-parallel (case 1) and the other is that they are parallel (case 2). Schematic pictures of these two cases are shown in Fig \ref{fig1}. Based on the standard magnetic dipole structure, the magnetic field lines from both NSs interact at a distance $r_i=a/(1+\epsilon^{1/3})$, where $\mu_*r_i^{-3}=\mu_{\rm c}(a-r_i)^{-3}$ and $\epsilon=\mu_{\rm c}/\mu_*$. For the anti-parallel case (case 1), the directions of magnetic field lines are opposite at $r_i$, and thus magnetic reconnection can happen. The total dissipated energy can be calculated as $B(r_i)^2V/8\pi$, where $B(r_i)=\mu_*r_i^{-3}$ is the magnetic field strength at $r_i$, and $V$ is the volume. In an orbital period ($T_{\rm o}=2\pi/\Omega$), the volume can be calculated as $V\approx2\pi r_i \dot{r_i}T_{\rm o} h$, where $h\approx0.77 r_i$ (see the Appendix A) is the possible highest height of the reconnection zone in a dipole magnetosphere as shown in Fig \ref{fig1}. Then we obtain
\be
L_{a,{\rm rec}}\approx B(r_i)^2V/8\pi T_{\rm o}\approx 1.5\times10^{43}
\mu_{*,30}^2(1+\epsilon^{1/3})^3 (a/ 30\;\!{\rm km})^{-7}{\rm erg/s}    \  ,
\ee
where $M_*=M_{\rm c}=1.4M_\odot$, and $\dot a=-{64\;\!G^3M_*^3q(1+q)/ 5 c^5a^3}$ \citep{Peters1964}.

After the magnetic reconnection, a UI can also form.
The electromotive force (EMF) in this case is $\cE \simeq l|{\bf E}|$, where
${\bf E}= \textbf{v}\times {\bf B_{\rm c}}/c$,
${\bf v=(\Omega-\Omega_*)}\times \textbf{\emph{a}}$,
and $l$ is the length where EMF generates.
In the polar coordinate system, the magnetic field line equation is $r=
r_{\rm max}\sin^2\theta$, where $r_{\rm max}$ is the maximum distance between
the field line and the NS. For the last close magnetic field line, we have
$r_{\rm max}=R_{\rm L}=c/\Omega_{\rm c}$, while for the last interacting magnetic
field line, we obtain $r_{\rm max}=a-r_i$ or $r_i$ for the companion or main star,
respectively. To calculate the angle between the magnetic field line and the magnetic axis, we take
$r=R_{\rm c}$. Then the length can be calculated as
$l=R_{\rm c} (\sin\theta(a-r_i)-\sin\theta(R_{\rm L}))$,
    where $\sin\theta(r_i)=(R_{\rm c}/a-r_i)^{1/2}$ is the angle between magnetic axis
      and the last interacting magnetic field line (see Fig. 1a),
      and $\sin\theta(R_{\rm L})=R_{\rm c}\Omega_{\rm c}/c$
is the angle between the magnetic axis and the last close magnetic field line.
Let's assume that $\Omega_*$, $\Omega_{\rm c}$ are small enough to be neglected.
This is appropriate, as the tidal torque cannot lead to spin-orbit synchronization during the inspiral \citep[e.g.][]{Kochanek1992,Bildsten1992,Lai1994,Ho1999}.
The resistance is usually considered to be dominated by the magnetosphere
\citep[$\cR=4*4\pi/c$,][]{Piro2012,Lai2012}.
We then obtain the energy-dissipation rate,
\be
L_{a,{\rm UI}}=2\cE^2/\cR \approx 3.8\times 10^{44} (R_{\rm c}/10\;\!{\rm km})^{-3}
  (1+\epsilon^{1/3})\epsilon^{5/3}\mu_{*,30}^2(a/30\;\!{\rm km})^{-2}
{\rm erg/s}  \   .\label{eq_a}
\ee

For the parallel case (case 2 in Fig. 1b), the magnetic field lines beyond $r_i$ will be
compacted into the area near $r_i$, thus the magnetic energy is stored.
Considering that the magnetic flux parallel to
$\bmu_*$ is conserved, we can obtain the compacted mean magnetic field
$B'$ around $r_i$ in the magnetosphere of the main star from
\be
\int^{\infty}_{r_i} B(r)r\;\!{\rm d}r=B'(r_i) r_i\;\! \delta r_i    \  ,
\ee
where $\delta r_i$ is the thickness of the compacted region. If we assume
$\delta r_i=\eta r_i$, we obtain $B'_*(r_i)=\mu_*\eta^{-1} r_i^{-3}$.
The energy stored in an orbital period is then $B'_*(r_i)^2 2\pi r_ih\eta r_i-
\int^{\infty}_{r_i} B^2(r)2\pi r^2{\rm d}r=(0.19/\eta-0.08)\mu_*^2 r_i^{-3}$.
Replacing $\mu_*$ and $r_i$ with $\mu_{\rm c}$ and $a-r_i$ respectively,
we obtain the stored magnetic energy of the companion star.
The total energy-dissipation rate is the sum of the dissipation rates by two NSs,
\ba
&L_{\rm p}&\approx(0.19/\eta-0.08)\mu_*^2 r_i^{-3}(1+\epsilon)/T_{\rm o} \\  \nonumber
&&\approx1.8\times 10^{43}(0.19/\eta-0.08)
\mu_{*,30}^2(1+\epsilon^{1/3})^3 (1+\epsilon)(a/ 30\;\!{\rm km})^{-9/2}{\rm erg/s}   \ ,   \label{eq_p}
\ea
  where $h=0.77\;\!r_i$ is also used.
In comparison with that in case 0,
   the energy-dissipation rates in case 1 and 2 are much higher.
Assuming that $B_*=B_{\rm c}=10^{12}$\,G, $\eta=0.1$, $R_*=R_{\rm c}=10\;\!{\rm km}$,
and setting $t=10$\,s when $a=R_{\rm c}+R_*$, we obtain the time evolution
of the energy loss rates for the three cases, as shown in Fig. 2. We find that the energy loss rates and their dependences on $a$ in our analytical calculations are in good agreement with the results from simulations \citep{Palenzuela2013b,Ponce2014}. 
We find that $L_{a,{\rm UI}}>10^{43}$\,erg/s, and $L_{\rm p}>10^{40}$\,erg/s for more than 10\,s,
which make them more probable to be observed. 

We now consider the cases for NS-WD binaries and double WD binaries.
Since we mainly consider the electromagnetic energy loss without significant
mass transfers, the separation of the two stars should be larger than the
tidal radius $a>r_t\approx q^{-1/3}R_{\rm c}$; for a typical WD, we have $M_{\rm c}\approx M_{\odot}$ and $R_{\rm c}=0.01R_{\odot}$.
The main star extends its magnetic field on the WD companion at a strength
$\mu_*/R_{\rm c}^3\approx2.1\times10^3\mu_{*,30}$\,G,
  while a moderately magnetised WD could easily have a magnetic field
   $B_{\rm c}\sim 10^6$\,G \citep{Norton1989,Wu1991}.
Therefore, the energy
dissipation should be calculated using the formulae in cases 1 and 2 of double NS binaries,
\ba
&L_{a,{\rm UI}}&\approx 1.1\times 10^{30} (R_{\rm c}/0.01R_{\odot})^{-3}(1+\epsilon^{1/3})
\epsilon^{5/3}\mu_{*,30}^2(a/3\times 10^4{\rm km})^{-2}
{\rm erg/s}  \  .\\
 &L_{\rm p}&\approx5.7\times 10^{29}(0.19/\eta-0.08)
\mu_{*,30}^2(1+\epsilon^{1/3})^3(1+\epsilon) (a/ 3\times 10^4\;\!{\rm km})^{-9/2}{\rm erg/s}   \   .
\ea
Note that WDs could have magnetic moments higher than those of the NSs.
It is known that WDs in close binaries, e.g. magnetic cataclysmic variables,
  have magnetic moment $\mu\sim10^{33}-10^{34}\;\!{\rm G\;\!cm}^3$ or even higher
  \citep[see][]{Wu1991}.

\section{Spectra of the EM counterparts}

\subsection{Photon spectra in double NS binary systems}

To derive the photon spectrum, we must first know the electron spectrum. Note in this paper, we do not distinguish electrons from positrons. For all three cases, the electric fields are generated in directions perpendicular to the magnetic fields, as will be shown below.
Therefore, the accelerated electrons suffer from synchrotron radiation cooling. 
Here we neglect the effects of the ${\bf E\times B}$ drift for simplification. Because NSs and WDs usually have very intense magnetic fields, we consider the synchrotron radiation in the quantum electrodynamics (QED) regime. The power spectrum for an electron with energy $\epsilon_e=\gamma m_e c^2$ is given by
\be
P_{\rm syn}(\omega)={ e^2\omega\over3^{1/2}\pi c \gamma^2}
{\int_y^{\infty}K_{5/3}(x){\rm d}x+{\hbar^2\omega^2\over m_e^2c^4
\gamma (\gamma-\hbar\omega/m_ec^2)}K_{2/3}(y)} \label{synQED}\ ,
\ee
 \citep{Akhiezer1994,Baring1988,Anguelov1999},  where
\be
y={2\hbar\omega B_{\rm cri}\over3 B\gamma(\gamma-\hbar\omega/m_ec^2)}   \   ,
\ee
and $B_{\rm cri}=m_e^2c^3/e\hbar=4.41\times 10^{13}$\,G is the critical measurement in the QED regime.
In the regime of $\hbar\omega\ll \gamma m_e c^2$, Eq. \ref{synQED} equals to the classical synchrotron radiation.

The maximum Lorentz factor can then be obtained by balancing the acceleration
with the synchrotron radiation cooling,
\be
\int P(\omega){\rm d}\omega=P_{\rm acc}\approx eEc    \  .    \label{eqgamax0}
\ee
Let us start with case 0, in which we have $E\approx \Omega \mu_* a^{-2}/c$. Using the classical synchrotron radiation formula $P_{\rm syn, tot}=2e^4B^2\g^2/3m_e^2c^3$, the maximum accelerated Lorentz factor is obtained as
$\gamma_{\rm max, acc}\approx 3.1\times10^2\mu_{*,30}^{-1/2}(a/ 30\;\!{\rm km})^{5/4}$ \citep{Wang2016}.
When $a<100$\, km, we should consider the QED modification, and the maximum accelerated Lorentz factor will be roughly two times larger.

The high-energy photons by synchrotron radiation may be absorbed by the magnetic field to produce electron-positron pairs in the magnetosphere.
The mean free path of a photon with energy $\hbar \omega$ in a magnetic field $B$ is
\be
\lambda=2.3\times10^{-8}B_{\rm cri}/B\exp{(8/3\chi)}~{\rm cm}\label{eq:mfp_g}\ ,
\ee
 \citep{Erber1966}, where $\chi=\hbar\omega B/m_ec^2B_{\rm cri}$. If we take $\lambda\sim R_{\rm c}$ for case 0, we obtain $\hbar\omega_{\rm max}\sim 1.3\times10^2\mu_{*,30}^{-1}(a/ 30\;\!{\rm km})^{3} m_ec^2$.
It should be noticed that $\hbar\omega_{\rm max}/m_ec^2>2$ is required to produce an electron-positron pair. 
Therefore, the photons emitted by the electrons with maximum accelerated Lorentz factor $\gamma_{\rm max, acc}$ could be absorbed. Taking this into consideration, we set $\gamma_{\rm max, syn}\leq \gamma_{\rm max, acc}$ as the maximum Lorentz factor for synchrotron radiation to make sure that the synchrotron photons will not exceed the absorption limit. 
 

We here define three partition parameters, $\eta_{\rm syn}$, $\eta_{\rm cur}$, and $\eta_{\rm BB}$, to represent fractions of the total energy dissipated by synchrotron, curvature, and blackbody radiation, respectively. Next, we consider curvature radiation. Its spectrum is analogous to classical synchrotron radiation with the Larmor radius being replaced by the curvature radius of the field line. The cooling time due to curvature radiation is
\be
t_{\rm cur}=\gamma m_ec^2/P_{\rm cur}=1.78\times 10^8 r_{\rm cur,6}^2\gamma_2^{-3}~{\rm s}   \  ,
\ee
where $r_{\rm cur,6}$ is the curvature radius in units of $10^6$\,cm and
$\gamma_2=\gamma/10^2$. Thus, if the curvature radiation is non-coherent, the electron will only lose a very small fraction of its energy to cross the magnetosphere within a time $t_{\rm cro}\sim a/c\sim 10^{-4}(a/30\;\!{\rm km})$\,s, namely $\eta_{\rm cur}\ll1$.
But if the curvature radiation is coherent, a FRB would be produced as studied in \cite{Wang2016}; in this case, $\eta_{\rm cur}$ can be much larger. However, we should note that only photons with frequency larger than the plasma cut-off frequency can escape from the magnetosphere \citep{Lyubarskii1998}, i.e. $\omega>\g^{-1/2}\omega_p=\left(4\pi n_e e^2/\g m_e\right)^{1/2}$. If we assume that the plasma density is of order of the density to screen the electric field, which is analogous to the Goldreich-Julian density $n_{e}\sim \Omega B_a/2\pi e c$ \citep{GJ1969}, the cut-off frequency is $\nu_p=\omega_p/2\pi\sim 1.8B_{a,11}^{1/2}\g_2^{-1/2} (a/ 30\;\!{\rm km})^{-3/4}$ GHz, where $B_a=10^{11}B_{a,11}$ G is the magnetic field of the acceleration region. This is consistent with the observation of FRBs, as FRBs are generally observed with frequencies larger than GHz.
The rest energy of the electrons will be dissipated through blackbody radiation after these electrons hit and heat the main NS's surface. Because the thermal conductivity in the direction paralleling to the magnetic field is much larger than that in the perpendicular direction \citep{Greenstein1983,Page1995,Geppert2004,Geppert2006}, the heat conduction happens only in the direction along the magnetic field lines. Therefore, we assume the blackbody radiation will be dominated by two confined spots with an area $S_{\rm *,BB}=2\pi R_*^2\sin^2\theta(a)=2\pi R_*^3/a$ (see Fig.\,\ref{fig1}a),
\be
L_{\rm BB}=L_{\rm *,BB}\approx S \sigma_{\rm SB} T_{\rm *,BB}^4=\eta_{\rm BB}L_{\rm UI}  \   ,
\ee
where $\sigma_{\rm SB}=5.67\times10^{-5}$\,erg cm$^{-2}$ s$^{-1}$ K$^{-4}$ is the Stefan-Boltzmann constant. The temperature of the hot spots is then $T_{\rm *,BB}=3.4\times10^8\eta_{\rm BB}^{1/4}B_{*,12}^{1/2} (a/ 30\;\!{\rm km})^{-3/2}$\,K. This is larger than the temperature induced by tidal heating, which can heat the NS up to $10^8$\,K before the final merge \citep{Lai1994}. Therefore we will neglect the tidal heating in this paper.

Using the same method, we can also calculate the electric field
$E$, the maximum accelerated Lorentz factor $\gamma_{\rm max, acc}$,
the maximum escaping photon energy $\hbar\omega_{\rm max}$, the
temperature $T_{\rm *,BB}$, and the area of the hot spots $S_{\rm *,BB}$
in cases 1 or 2. We summarize them in Table \ref{tab1}. The electric field
generated in case 2 is
\be
{E\over\delta r_i }\approx{B'(r_i)-B(r_i)\over c\;\!\delta t}\     ,
\ee
which is different from those in cases 0 and 1. If we assume $\delta r_i/\delta t \sim r_i \Omega$, then $E\sim\mu_{*}r_i^{-3}(\eta^{-1}-1)a \Omega/c$. Interestingly, there are not only two hot spots on the main star surface, but
also two on the companion star surface with temperatures $T_{\rm c,BB}=\epsilon^{1/4}(R_{\rm c}/R_*)^{-3/4}T_{\rm *,BB}$. The total blackbody radiation luminosity is then $L_{\rm BB}=L_{\rm *,BB}+L_{\rm c,BB}$.
We show the typical spectra of different cases at $1$\,s before the NSs come into contact in Fig. \ref{fig3} with an assumption of $B_*=10^{12}$\,G and Fig. \ref{fig4} with an assumption of $B_*=10^{10}$\,G. The electron spectrum for calculating the synchrotron radiation is assumed to be $dN/d\gamma\propto \gamma^{-2}$ with a minimum Lorentz factor $3$ for a relativistic electron. This is the typical electron spectrum induced by the synchrotron-pair cascades without injections from other sources \citep{Wang2018}. The synchrotron spectrum is calculated using Eq. \ref{synQED}. We should note that the observed spectra might differ a bit from these theoretical ones. A detailed discussion about this is in section \ref{obs}.

We now calculate the possible screening effect due to the generation of pairs, which happens in the polar caps of pulsars \citep[e.g.][]{Harding1998}. Considering that product of the size and the magnetic field of the acceleration region is $\sim 10^{17}$ cm$\cdot$G, the maximum escape photon energy will be around $\chi\sim 0.1$ \citep[][see also Eq. \ref{eq:mfp_g}]{Wang2018}. These photons are produced by electrons with energy $\g B_a/B_{\rm cri}\sim 1$ (based on Eq. \ref{synQED}), namely, $\g\sim4.4\times10^2B_{a,11}$. This is almost the maximum accelerated Lorentz factor. Despite the screening, the results that we have obtained above still hold.

\begin{table}
\centering
\begin{tabular}{|l|m{1.8cm}|m{4.2cm}|m{3.cm}|m{3.5cm}|m{1.7cm}|}
\hline
Cases & $E$ & $\gamma_{\rm max, acc}~(10^2~{\rm for~electron},$\newline$3.37\times10^8\,{\rm for~ proton})$ & $\hbar\omega_{\rm max}~(m_ec^2)^a$ & $T_{\rm *,BB}~(10^8$K)& $S_{\rm *,BB}$\\ \hline
Case 0 & $\Omega \mu_* a^{-2}c^{-1}$ &$6.2 \mu_{*,30}^{-1/2}(a/ 30\;\!{\rm km})^{5/4}$& $1.3\times10^2\mu_{*,30}^{-1}$\newline$
(a/ 30\;\!{\rm km})^{3}$& $3.4\eta_{\rm BB}^{1/4}
B_{*,12}^{1/2}$\newline$ (a/ 30\;\!{\rm km})^{-3/2}$&$2\pi R_*^3a^{-1}$\\ \hline
Case 1 & $a\Omega B_{\rm c}c^{-1}$ & $ 2.0B_{{\rm c},12}^{-1/2} (a/ 30\;\!{\rm km})^{-1/4}$&$4.3B_{{\rm c},12}^{-1}$\newline$\ln^{-1}(0.98B_{{\rm c},12})$&$13\eta_{\rm BB}^{1/4}
B_{*,12}^{1/2} $\newline$\epsilon^{5/12} (a/ 30\;\!{\rm km})^{-1/4}$&$2\pi R_*^3a^{-1}$\newline$(1+\epsilon^{1/3})$\\\hline
Case 2 $^b$ & $\mu_{*}r_i^{-2}\Omega c^{-1}$\newline$(\eta^{-1}-1)$&$6.2(\eta^{-1}-1)^{-1/2
}B_{*,12}^{-1/2}$\newline$(1+\epsilon^{1/3})^{-2} (a/ 30\;\!{\rm km})^{5/4}$&$8.1
B_{*,12}^{-1}(a/ 30\;\!{\rm km})^{3}$\newline$ (1+\epsilon^{1/3})^{-3}$&$4.4(a/ 30\;\!{\rm km})^{-7/8}\eta_{\rm BB}^{1/4}$
\newline$\epsilon^{1/12}
(1+\epsilon^{1/3})^{1/4}$\newline$ B_{*,12}^{1/2}(0.19/\eta-0.08)^{1/4}$&$2\pi R_*^3a^{-1}
$\newline$(1+\epsilon^{1/3})^2$\newline$\epsilon^{-1/3}$\\
\hline
\end{tabular}
\caption{$^a$ The minimum value required to produce an electron-positron pair is $\hbar\omega_{\rm max}=2m_ec^2$; and the corresponding electron's Lorentz factor is $\gamma_{\rm max, syn}\approx (\omega_{\rm max} m_e c/0.44 e B)^{1/2}$.
$^b$ In case 2, $\gamma_{\rm max, acc}$ is roughly 2 times larger when $a<200$\,km.
The maximum escaped photon energy is calculated by $\lambda\sim \eta r_i$. And the
area of the hot spots is calculated with an assumption $R_*=R_{\rm c}$. \label{tab1}}
\end{table}

\subsection{Photon spectra in the NS-WD or double WD binary system}

In NS(WD)-WD binary systems, the classical synchrotron radiation formula is appropriate. In case 1, an electron is accelerated on the WD surface with a maximum Lorentz factor
$\gamma_{\rm max, acc}=7.9\times 10^3(a/3\times 10^4\;\!{\rm km})^{-1/4}B_{\rm c,6}^{-1/2}$.
Two hot spots form on the main compact star with a typical the temperature
$T_{\rm *,BB}=1.7\times10^6\eta_{\rm BB}^{1/4} \mu_{*,30}^{1/2}\epsilon^{5/12} (a/ 3\times10^4\;\! {\rm km})^{-1/4}(R_{\rm c}/0.01R_{\odot})^{-3/4}(R_*/10\;\! {\rm km})^{-3/4}$\,K.
In case 2, the maximum Lorentz factor is $\gamma_{\rm max, acc}
=1.7\times 10^6(\eta^{-1}-1)^{-1/2}(1+\epsilon^{1/3})^{-2} (a/3\times 10^4\;\!{\rm km})^{5/4}\mu_{*,30}^{-1/2}$.
The temperature of the hot spots on the main star is $T_{\rm *,BB}=1.5\times10^6\eta_{\rm BB}^{1/4}\mu_{*,30}^{1/2}(1+\epsilon^{1/3})^{1/2} (a/ 3\times10^4 {\rm km})^{-7/8}(R_*/10\;\! {\rm km})^{-3/4}(0.19/\eta-0.08)^{1/4}$\,K, and the temperature of the hot spots on the companion is $T_{\rm c,BB}=\epsilon^{1/4}(R_{\rm c}/R_*)^{-3/4}T_{\rm *,BB}$. We show the typical spectra in a NS-WD binary and a double WD binary in Fig.\,\ref{fig5} and \ref{fig6}, respectively.
It should be noted that there are four hot spots for the parallel cases, and the temperature on the WD is quite different with the temperature on the NS.
Thus there are two blackbody components in Fig.\,\ref{fig5}.

\subsection{Acceleration of protons}

In these binary systems, protons in the magnetosphere are also accelerated by the electric field.
Replacing the electron mass, $m_e$ with proton mass, $m_p$ in the expression for synchrotron radiation,
  gives a maximum Lorentz factor for the proton  $(m_p/m_e)^2=3.37\times 10^6$ times
  larger than the $\gamma_{\rm max, acc}$ of the electrons.
It is therefore worthy to assess how close this maximum accelerated Lorentz factor would be achieved.
The maximum energy that an electric field accelerator can produce is $\epsilon_{\rm max, E}= eEl$.
This is practically what we would expect if simply taking the \cite{Hillas1984} criterion, $\epsilon_{\rm max, acc}\leq e B l$,
  which is a sensible approximation, as the electric field is always smaller than the magnetic field, $E\sim B a\Omega/c<B$
  in the model configuration of the system adopted in this study.
We calculate this maximum energy for each case:
  $\epsilon_{\rm max, E}=4.1\times10^{18}\mu_{*,30}(a/ 30\;\!{\rm km})^{-7/2}$\,eV for case 0,
  $\epsilon_{\rm max, E}=6.4\times10^{19} B_{{\rm c},12}(a/ 30\;\!{\rm km})(1+\epsilon^{-1/3})^{1/2}$\,eV for case 1,
  and $\epsilon_{\rm max, E}=1.2\times10^{19}\mu_{*,30}(1+\epsilon^{1/3})\eta(\eta^{-1}-1)$\,eV for case 2.
 As shown in column 3 of Table\,\ref{tab1}, the maximum accelerated proton energy in all three cases is around $\epsilon_{\rm max}=3.37\times10^6 \gamma_{\rm max, acc}m_p c^2\sim6.4\times 10^{17}$\,eV, which does not exceed $\epsilon_{\rm max, E}$. Thus, the maximum proton Lorentz factor can be safely calculated by multiplying the maximum accelerated electron Lorentz factor by $(m_p/m_e)^2$. The partition parameter of charged particles, $\eta_{\rm CP}$ is expected to be very small, as the acceleration efficiency of electrons is much higher, and the fraction of protons should be far smaller than the fraction of electrons and positrons.

\section{Observational properties\label{obs}}

The partition parameters can affect the observed spectrum significantly. In the acceleration region, the energy is deposited into both the electrons and their synchrotron photons. For an electron accelerated to the energy $E_e=\gamma m_e c^2=eEct_{\rm acc}$ in a time $t_{\rm acc}$, the energy going into synchrotron radiation is $E_{\rm syn}=\int_0^{ t_{\rm acc}}P_{\rm syn, tot}{\rm d}t=2e^3B^2\g^3/9m_ec^2E$.
Therefore, the ratio of the kinetic energy of electrons to the total energy is $\eta_e=E_e/(E_e+E_{\rm syn})$.
For all three cases, we find $0.5<\eta_e<1$. If the velocity direction of these electrons is isotropic, around two-thirds of their energy will be emitted via the synchrotron radiation, based on the equipartition theorem. As a result, we have $\eta_{\rm syn}=1-\eta_e +2\eta_e/3 =1-\eta_e/3$, and $\eta_{\rm cur}+ \eta_{\rm BB}+\eta_{\rm CP}=\eta_e/3$. Therefore, the dominant component is the synchrotron radiation. This synchrotron radiation happens around the acceleration region in an opening angle depending on the electron velocity distribution, while the blackbody radiation is confined in the hot spots around the magnetic pole. The curvature radiation is beamed almost parallel to the magnetic field line. Because these three components (the blackbody, the synchrotron, and the curvature) may point to different directions, it is unlikely to observe all of them from the same source. Due to the orbital motion of the binary, these three components may behave periodically, but with a very short period (of the order ms in the late inspiral).

The EM signals from the inspiral of NS-NS(BH) binaries in the last few seconds can be responsible for the precursors of short GRBs. \cite{Troja2010} searched in 38 {\it Swift} short GRBs and found four precursor candidates, but with significance $<5.5\sigma$. \cite{Minaev2017} found only three candidates in 519 short GRBs detected by the SPI-ACS/INTEGRAL experiment, but with much higher significance. A precursor candidate is assumed to be weaker than the main burst, to antedate the main burst for more than 2\,s in \cite{Minaev2017}. They also analyzed the spectrum of the precursor for the individual bursts: GRB~090510
  \citep[discovered by][]{Troja2010}, and GRB~100717 \citep[see Table 3 in ][]{Minaev2017}. The time lag between the precursor and the main burst is 0.45\,s in GRB~090510, and 3.3\,s in GRB~100717. The time lag between GW170817 and GRB~170817A is 1.7\,s; therefore, we will only consider GRB~100717 in this paper. We note that this GRB is regarded as a long GRB in the Fermi catalog with $T_{90}=5.95\pm1.51$\,s \citep[see, for example in][]{Narayana_Bhat2016}, but in the SPI-ACS/INTEGRAL experiment, it's recognised as a short GRB, which is also supported by the behavior that there is no statistically significant spectral lag between the light curves in different energy ranges \citep[see more detailed discussions in][]{Minaev2017}. Therefore, we treat it as a short gamma ray burst with a precursor. Also the optimal spectral models for the precursor and main burst are found to be different between the precursor and the main burst \citep{Minaev2017}.

We analyze the precursor data of GRB~100717 in the energy range (8\,keV,  40\,MeV) detected by Fermi GBM.
Fig. \ref{fig7} is an example of light curve of GRB~100717, and the shaded region in the light curve is regarded as the precursor. Fig. \ref{fig8} is the spectrum of the precursor. We use four models in the official RMFIT software package to fit the spectrum: a simple power law, a power law with an exponential cut-off, a power law with an additional thermal component, and a power law with a break \citep[referring to as the Band function in][]{Band1993}. The power law model with/without an exponential cut-off in the energy range (8\,keV,  40\,MeV) corresponds to the synchrotron component in our model (see Fig. \ref{fig3} and \ref{fig4}), and the thermal component corresponds to the radiation from the hot spots. The Band function is a typical spectral type for the prompt emission of GRBs \citep{Band1993}. The optimal spectral energy distribution (SED) model is a power-law with a cut-off with a index $1.68\pm0.34$ and $E_{\rm peak}=984.8\pm420$\,keV, as shown in Fig \ref{fig8}, which hints that this precursor can be explained with the synchrotron radiation in our model. The spectral cutoff starts at $\sim1$\,MeV and means that the magnetic fields of the NSs are roughly $>10^{12}$\,G for different cases (see column 4 of Table 1). In this case, we roughly have $\gamma_{\rm max, syn}<\sim 30$, therefore it can be treated as monoenergetic electrons, and the low energy part of the synchrotron SED will behave as $\nu F\propto \nu^{4/3}$, which is consistent with the best-fitted SED power-law index $1.68\pm0.34$.

\section{Summary and discussion}
We have studied the EM energy-dissipation rate and the spectrum due to an interaction between the magnetospheres of double compact stars, based on the model similar to a UI. This process takes place once the magnetospheres of two compact stars come into contact. Even if the separation is larger than the light cylinder of the NS, this UI is likely to be established \citep{Mottez2011b,Mottez2011a,Mottez2014}, through the azimuthal magnetic fields in the pulsar winds from aligned pulsars \citep{Kirk2009}. In the late inspiral, the tidal deformation can become important. In the NS-WD binary, the WD can be disrupted and form a debris disk \citep[for example, see][]{Margalit2017,Fernandez2013,Bobrick2017,Zenati2018}. 
In this disruption stage, our model becomes invalid. In the eccentric or
hyperbolic NS-NS(BH) systems, fairly isotropic flares can be produced by the crust shattering \citep{Tsang2012,Tsang2013}. If the constraint on the NS equation of state by GW170817 is given \citep{Abbott2017a}, this crust shattering is likely to happen around $<1$\,s before the merger \citep{Tsang2012}. Our calculations may also be invalid in this case, as the magnetic fields of the magnetospheres will be significantly disrupted in the crust shattering. However, such eccentric or hyperbolic systems are very rare systems with an optimal occurrence rate $0.2-60$ Gpc$^{-3}$yr$^{-1}$ \citep{Tsang2013}, while the occurrence rate of the double NS merger is $1540^{+3200}_{-1220}$Gpc$^{-3}$yr$^{-1}$ \citep{Abbott2017a}. Therefore, our model holds for most double NS systems.

For double NS binaries, three cases are studied as examples: case 0 with $\mu_*\gg \mu_{\rm c}$, case 1 with $\bmu_*\sim-\bmu_{\rm c}$, and case 2 with $\bmu_*\sim \bmu_{\rm c}$. The EM energy-loss rates in cases 1 and 2 are much higher, and the dependence on separation $a$ are much weaker than that in case 0.
The high-energy photon spectra of these three cases consist of a characteristic blackbody radiation component and a synchrotron radiation component. At $\sim 1$\,s before the merger, the blackbody temperature peaks at around $11\eta_{\rm BB}^{1/4}B_{*,12}^{1/2}$\,keV$\sim78\eta_{\rm BB}^{1/4}B_{*,12}^{1/2}$\,keV. It should be noted that for case 2, there are two blackbody components from both stars. The synchrotron radiation components extend to MeV $\sim$ GeV, depending on the absorption limits of the magnetic fields. An FRB can be induced if the curvature radiation is coherent. Meanwhile, charged particles could also be accelerated in these systems with maximum energies around EeV. About 1 h before the merger, the temperature is around $3\eta_{\rm BB}^{1/4}B_{*,12}^{1/2}$\,keV$\sim45\eta_{\rm BB}^{1/4}B_{*,12}^{1/2}$\,keV, while the synchrotron radiation reaches from a few tens to a few hundreds MeV.
Similar calculations are performed to study the EM signals in NS-WD and double WD binaries. In these binaries at a separation $a=3\times10^4$\,km, the blackbody components peaks at around 0.1\,keV $\sim$ 10\,keV, while the synchrotron radiation component can reach to only a few MeV.

The partition parameters for different components are $\eta_{\rm syn}=1-\eta_e/3 $ and $\eta_{\rm cur}+ \eta_{\rm BB}+\eta_{\rm CP}=\eta_e/3$ with $0.5<\eta_e<1$. Therefore, the most significant component is the synchrotron radiation. We note that the spectrum in the BH-NS system is dominated by the curvature radiation \citep{D'Orazio2016}. This is different from our studies, since the electrons in the BH-NS systems are accelerated to a Lorentz factor $\sim10^7$ by the electric fields along the magnetic field line \citep{D'Orazio2016}. It should be noted that the blackbody component, synchrotron component, and possible FRB can point to different directions. As the short GRB and its afterglow are also beamed to a small angle, it will be very hard to observe all these EM counterparts for the same GW event. Thus, there is no surprise that no FRB is detected to be associated with GW170817 \citep[e.g.][]{Hallinan2017}.
Such an EM signal in the last few seconds pre-merger can be responsible for a precursor of the short GRB. We find that the spectrum of the precursor of GRB~100717 can be explained with the synchrotron component in our model, with an assumption that the magnetic fields of the NSs are $>10^{12}$\,G. Based on the sensitivity of Fermi GBM,\footnote{\url{https://fermi.gsfc.nasa.gov/science/instruments/table1-2.html}} the optimal detection limit is
$D_{\rm L}=100\;\!\xi B_{*,12}^2$\,Mpc,
where we assume $a=20$\,km and $\xi=0.02,~1.6,~0.7$ for case 0, 1, and 2 respectively.

\acknowledgements
We thank Prof. Xuefeng Wu and Dr. Yuanpei Yang for helpful discussions and an anonymous referee for helpful suggestions.
This work was supported by the National Basic Research Program of China (973 Program, Grant No. 2014CB845800),
the National Key Research and Development Program of China (Grant No. 2017YFA0402600)
and the National Natural Science Foundation of China (Grant No. 11573014). 
Fang-Kun Peng acknowledges support from the Doctoral Starting up Foundation of Guizhou Normal University 2017 (GZNUD[2017] 33).
KW thanks the hospitality of School of Astronomy and Space Science at Nanjing University during his visits.

\appendix
\section{The scale height of the interaction zone}
The magnetic field line equation for the magnetic dipole in the polar coordinates is $r= r_{\rm max}\sin^2\theta$, where we take $r_{\rm max}=r_i$ here. Therefore, in the Cartesian coordinates, we have $x=r\sin \theta=r^{3/2}r_i^{-1/2}$, and $y=r\cos \theta=r\sqrt{1-r/r_i}$,  where we only consider the two dimensional case and assume the magnetic dipole moment is parallel to the $y$-axis. Then we rewrite $y=y(x)$, and find that $dy/dx=0$ takes place at $(x_0,y_0)=(0.54r_i,0.39r_i)$, therefore, we take the scale height $h\approx2y_0=0.77r_i$.
\software{RMFIT (https://fermi.gsfc.nasa.gov/ssc/data/analysis/rmfit/)}
\bibliographystyle{apj}
\bibliography{ref}

\begin{figure}
\centering
\includegraphics[width=0.5\textwidth]{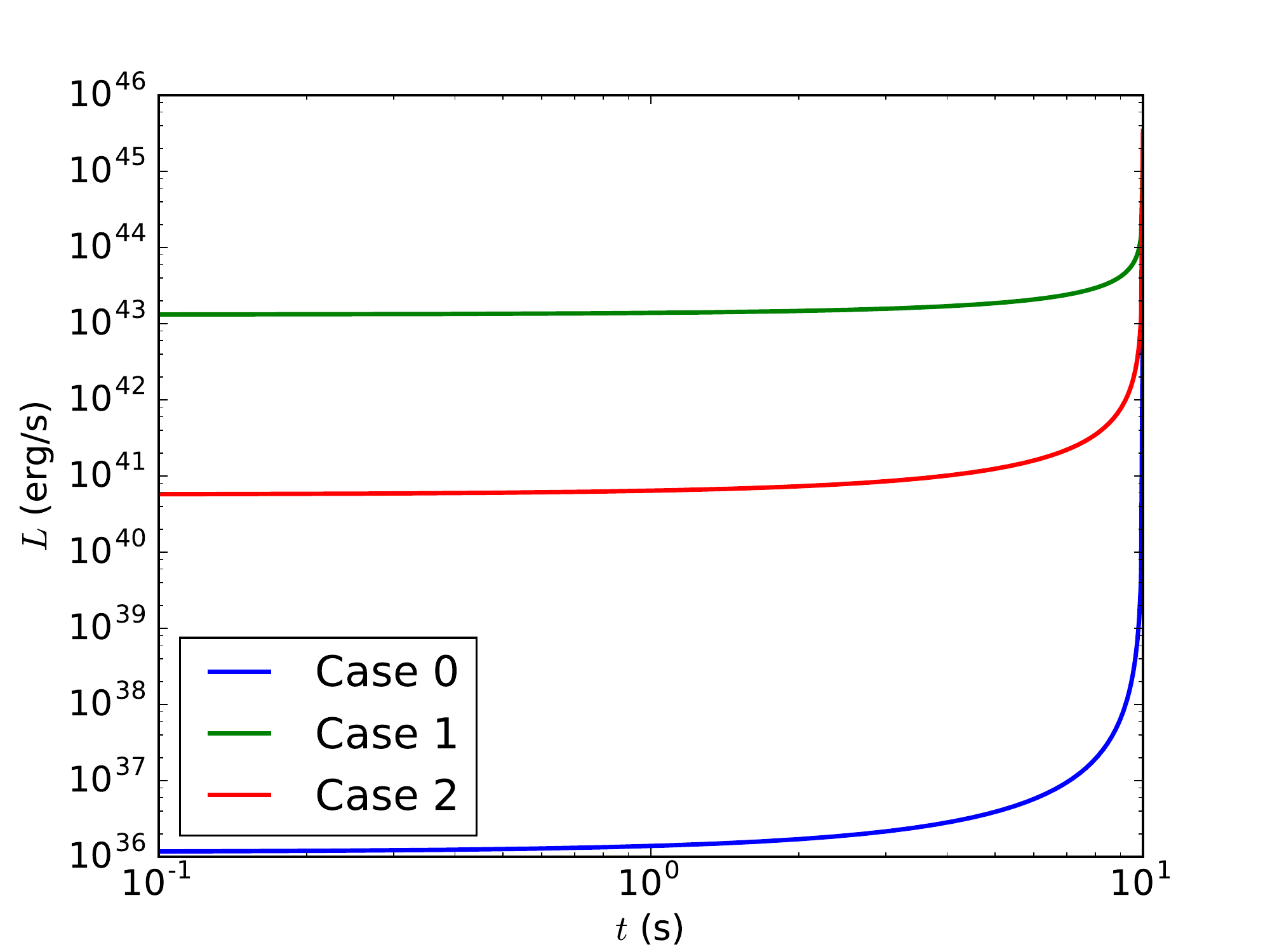}
\caption{Energy loss rates in three cases are shown, where we set $t=10$\,s
when two NSs come into contact. \label{fig2}}
\end{figure}

\begin{figure}
\centering
\includegraphics[width=0.5\textwidth]{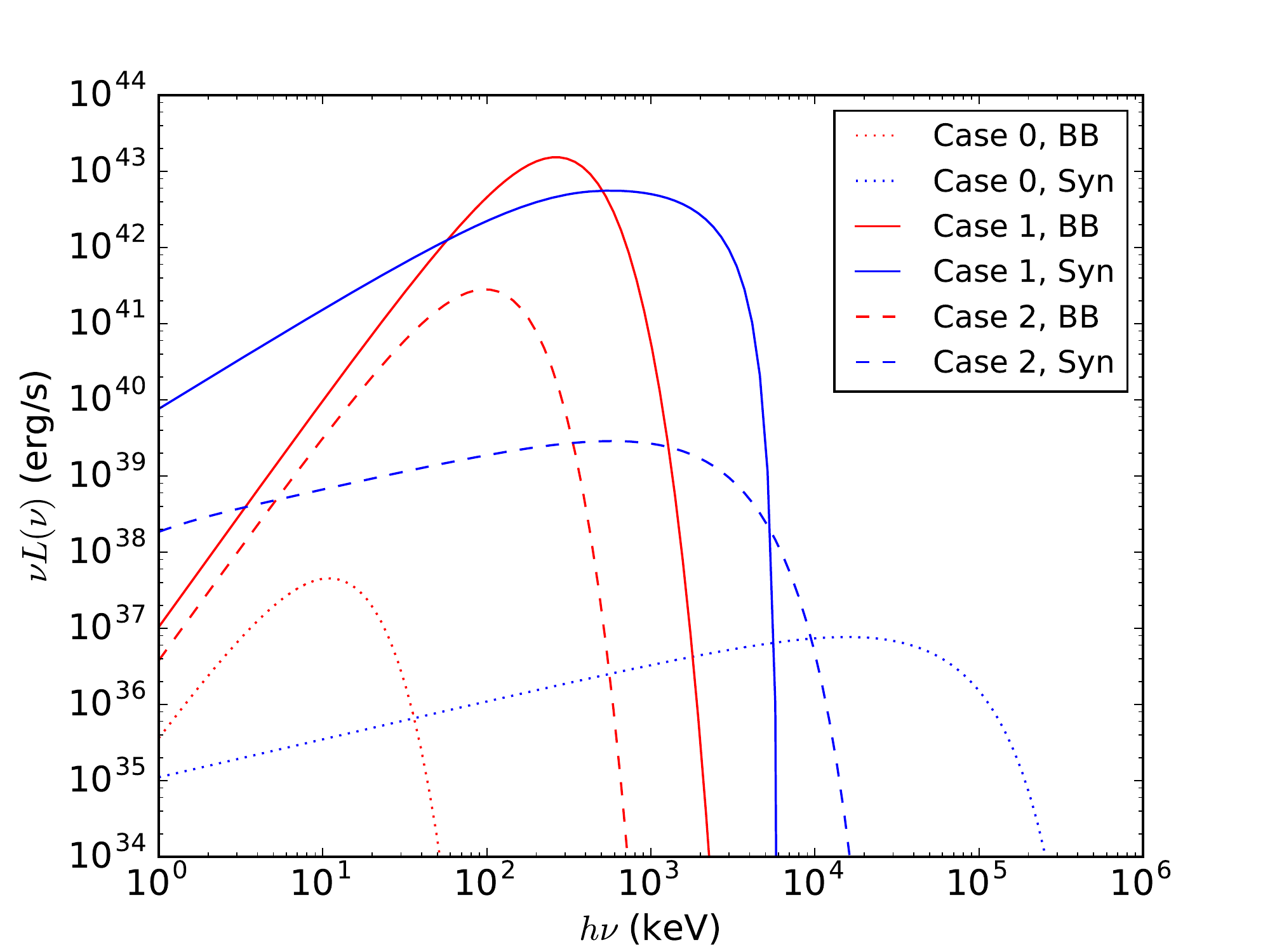}
\caption{$\nu L_{\nu}$ vs $h\nu$ plots of three cases at 1\,s before the
NSs come into contact are shown with different shapes of
lines. The red lines are the blackbody component, while the blue lines represent the
synchrotron component. The parameters of the binary systems are
chosen as $B_{*,12}=1$, $\epsilon=1$,
$\eta_{\rm syn}=0.5$, $\eta_{\rm BB}=0.5$ and $\eta=0.1$. \label{fig3}}
\end{figure}

\begin{figure}
\centering
\includegraphics[width=0.5\textwidth]{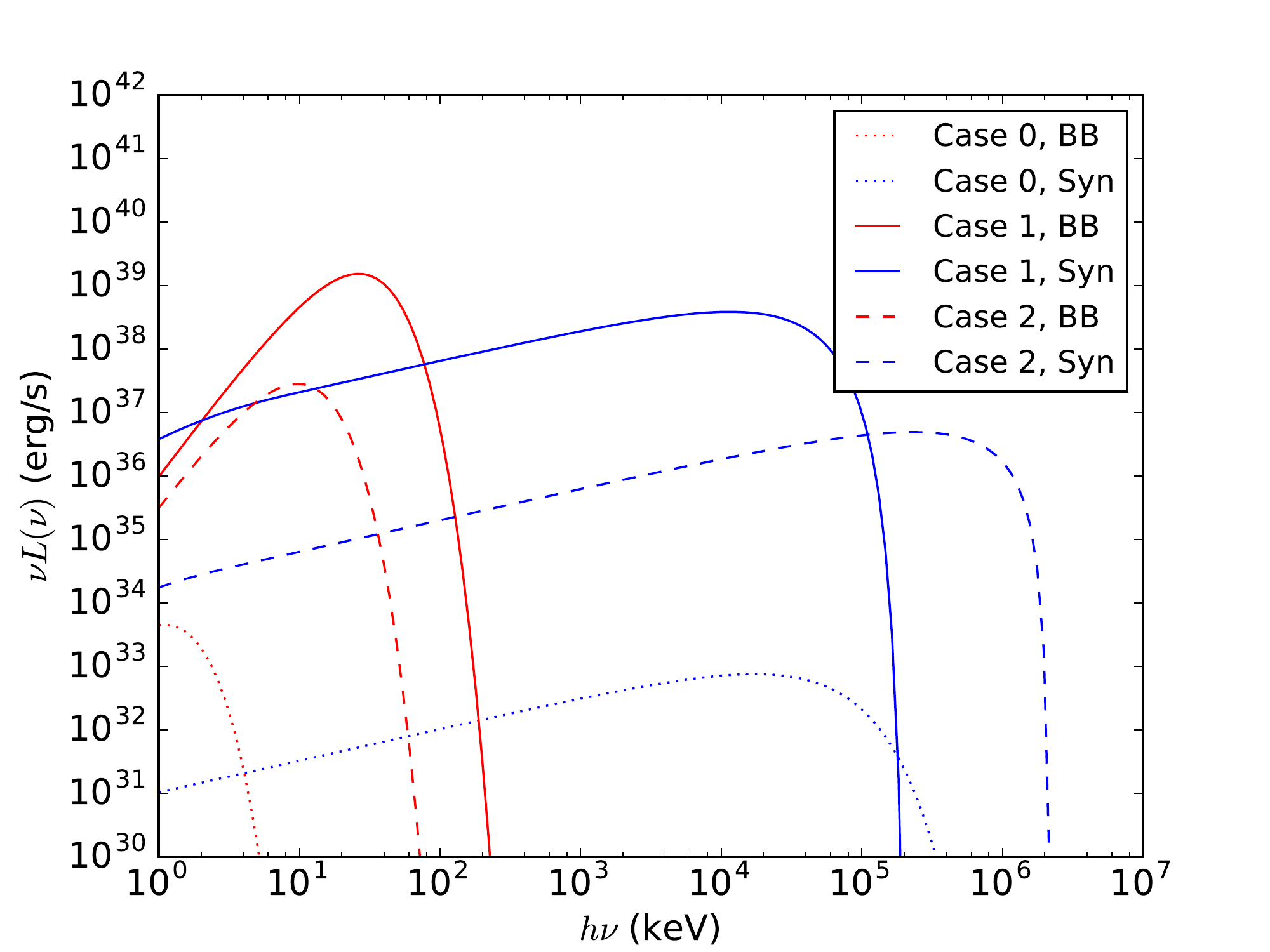}
\caption{$\nu L_{\nu}$ vs $h\nu$ plots of three cases at 1\,s before the
NSs come into contact are shown with different shapes of
lines. The red lines are the blackbody component, while the blue lines represent the
synchrotron component. The parameters of the binary systems are $B_{*,12}=0.01$,
$\epsilon=1$, $\eta_{\rm syn}=0.5$, $\eta_{\rm BB}=0.5$ and $\eta=0.1$. \label{fig4}}
\end{figure}

\begin{figure}
\centering
\includegraphics[width=0.5\textwidth]{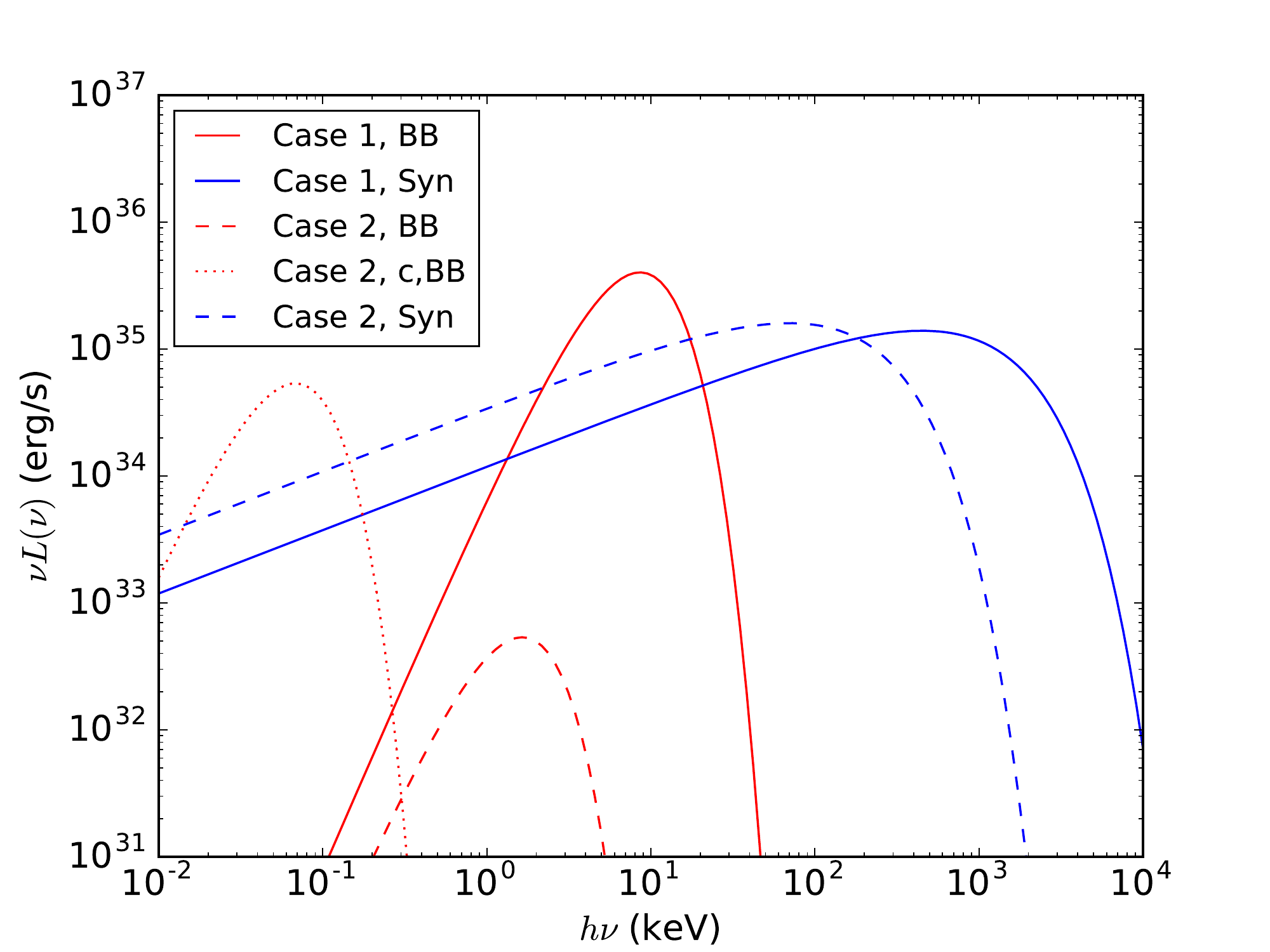}
\caption{$\nu L_{\nu}$ vs $h\nu$ plots of two cases in NS-WD binaries at
$a=3\times10^4$\,km are shown with different shapes of lines. The red lines
are the blackbody component, while the blue lines represent the synchrotron
component. The subscript $*$ and c represent the contribution of the
main star (NS) and the companion (WD), respectively.
The parameters of the binary systems are $\mu_{*,30}=1$, $\epsilon=1000$,
$\eta_{\rm syn}=0.5$, $\eta_{\rm BB}=0.5$ and $\eta=0.1$. \label{fig5}}
\end{figure}

\begin{figure}
\centering
\includegraphics[width=0.5\textwidth]{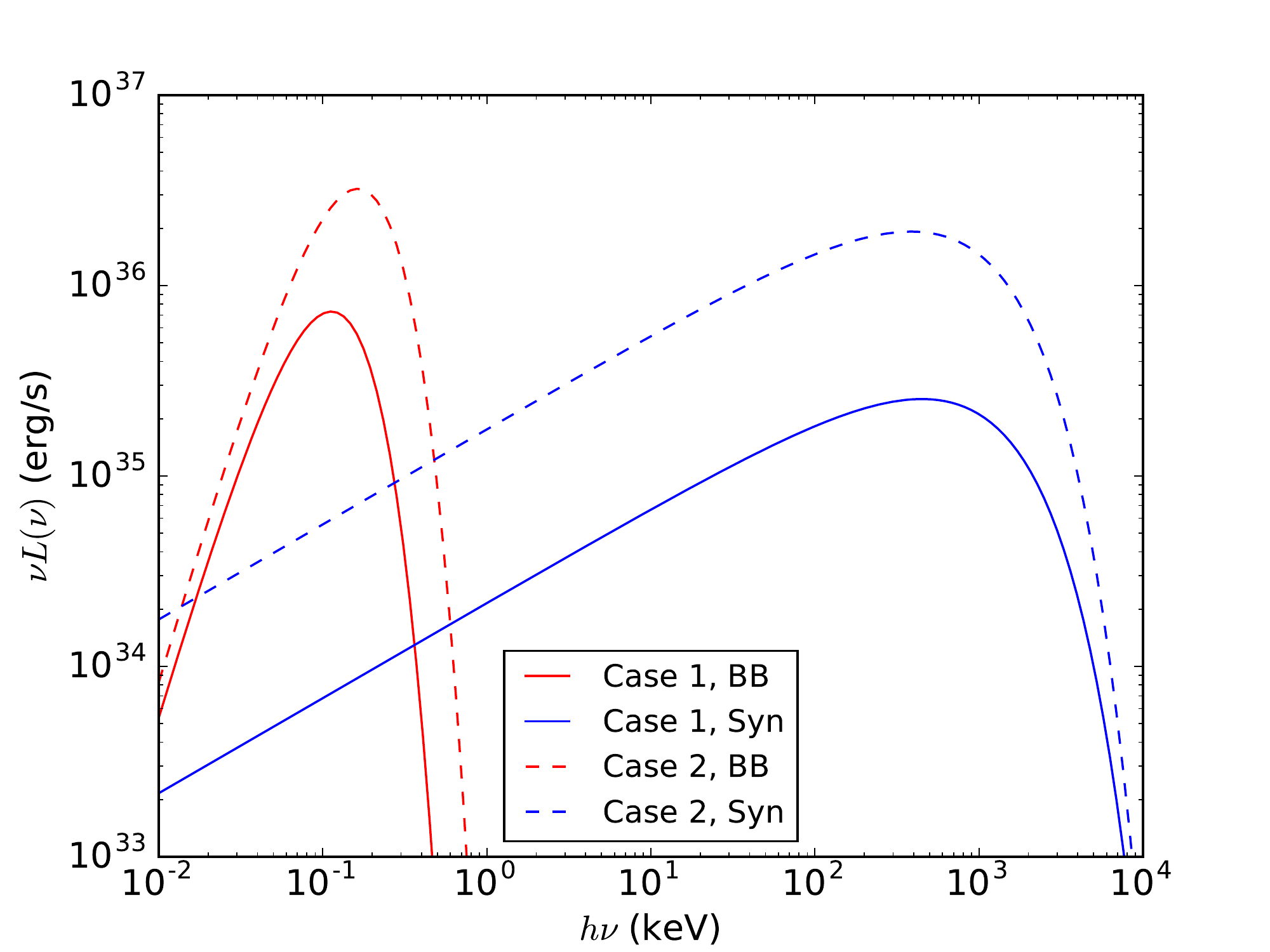}
\caption{$\nu L_{\nu}$ vs $h\nu$ plots of two cases in double WD binary at
$a=3\times10^4$\,km are shown with different shapes of lines. The red lines
are the blackbody component, while the blue lines represent the synchrotron
component. The parameters of the binary systems are $\mu_{*,30}=1000$, $\epsilon=1$,
$\eta_{\rm syn}=0.5$, $\eta_{\rm BB}=0.5$ and $\eta=0.1$. \label{fig6}}
\end{figure}

\begin{figure}
\centering
\includegraphics[width=0.5\textwidth]{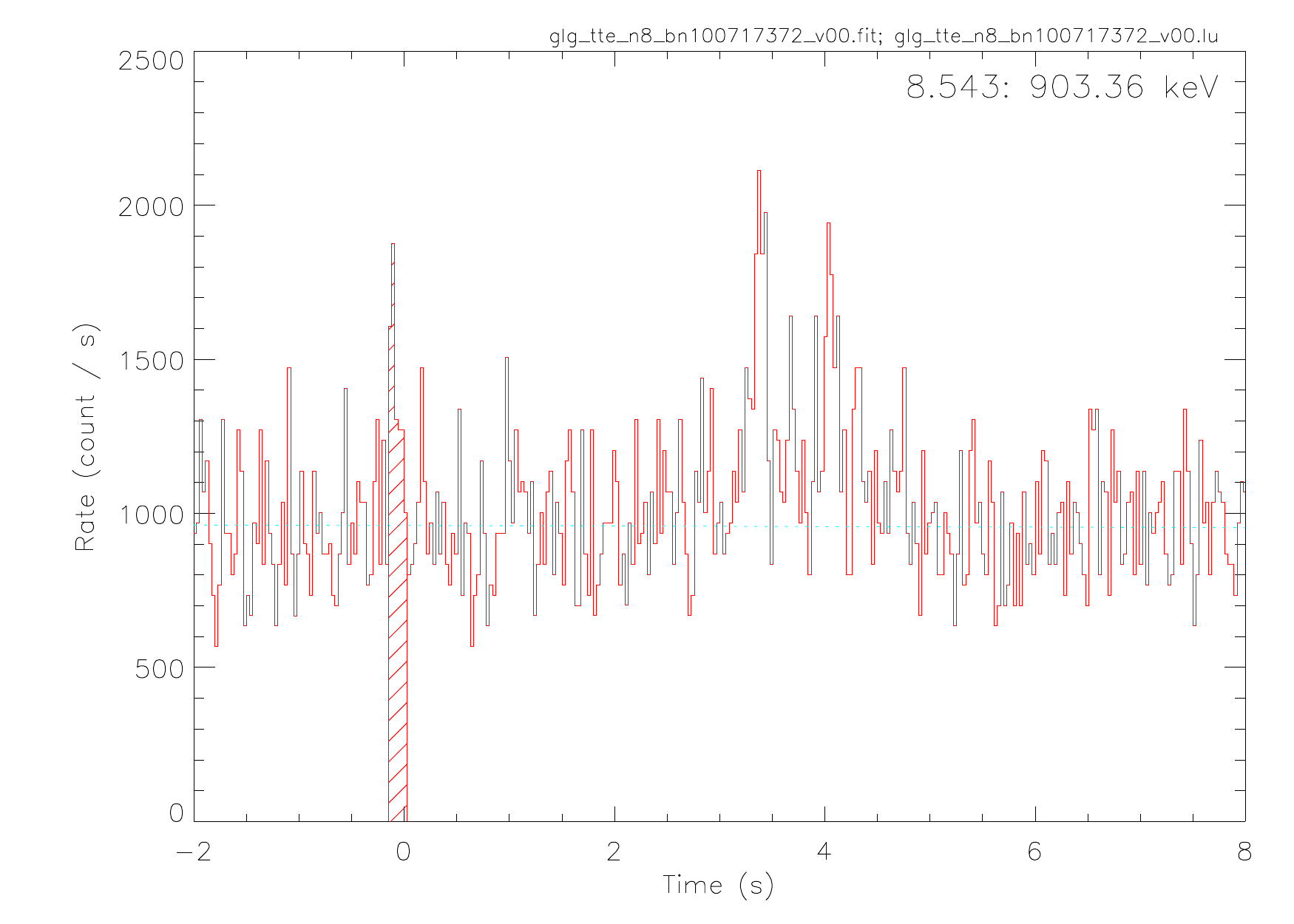}
\caption{The light curve of GRB 100717 detected by NaI08. The shaded region is treated with as the precursor.\label{fig7}}
\end{figure}

\begin{figure}
\centering
\includegraphics[width=0.5\textwidth]{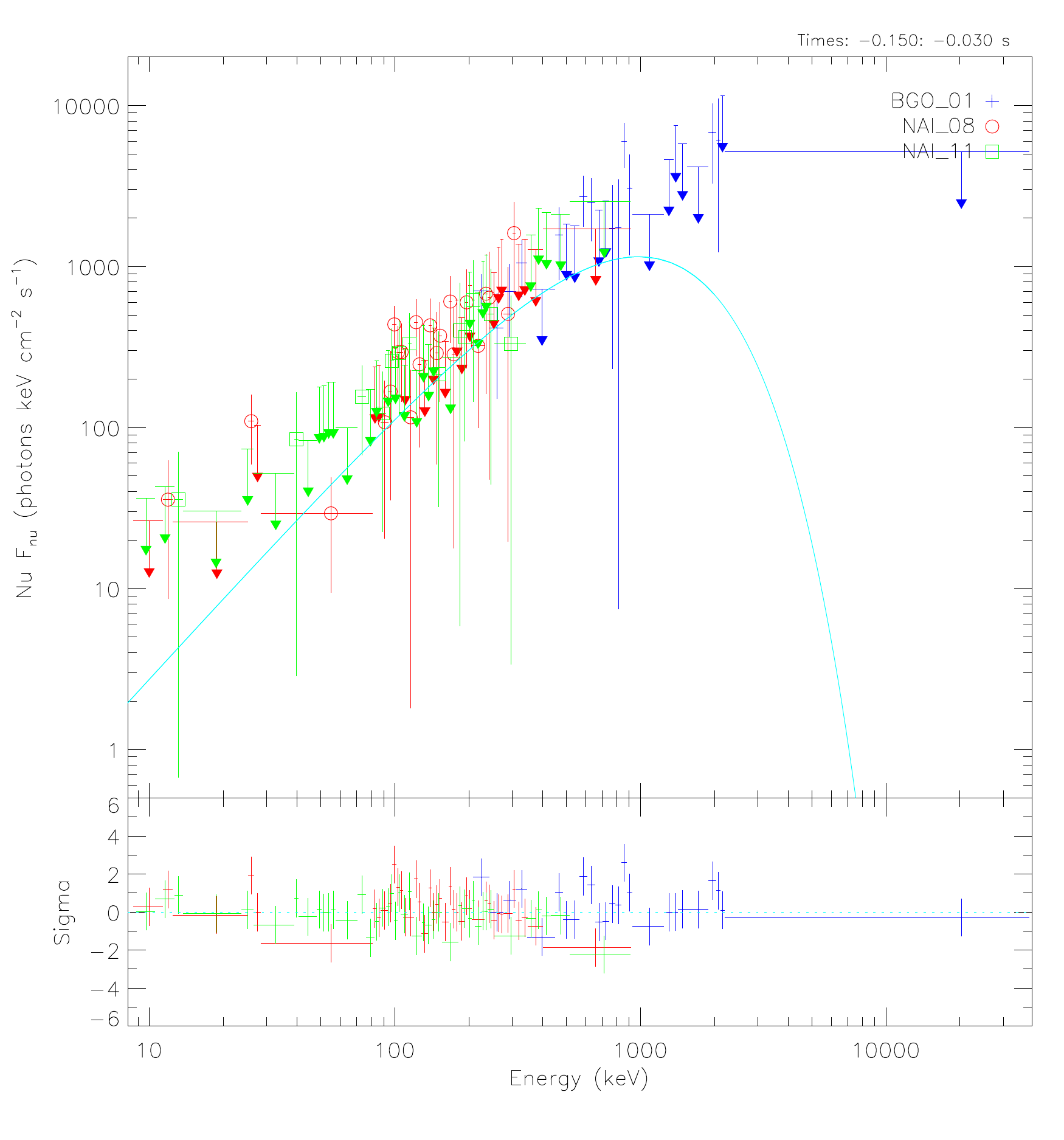}
\caption{The energy spectrum $\nu F_{\nu}$ of the precursor of GRB 100717 detected by NaI08, NaI11, and BGO01 detectors. The cyan curve is the optimal model.\label{fig8}}
\end{figure}

\end{document}